\documentclass[twocolumn,showpacs,preprintnumbers,amsmath,amssymb,aps]{revtex4}
\usepackage{graphicx}
\usepackage{dcolumn}
\usepackage{bm}
\usepackage{float}
\usepackage{soul}

\begin{document}

\title{Estimating the HF coupling parameters of the avian compass by comprehensively considering the available experimental results}

\author{Bao-Ming Xu}
\affiliation{School of Physics, Beijing Institute of Technology, Beijing 100081, People's Republic of China}

\author{Jian Zou}
\email{zoujian@bit.edu.cn}
\affiliation{School of Physics, Beijing Institute of Technology, Beijing 100081, People's Republic of China}

\author{Jun-Gang Li}

\affiliation{School of Physics, Beijing Institute of Technology, Beijing 100081, People's Republic of China}
\author{Bin Shao}

\affiliation{School of Physics, Beijing Institute of Technology, Beijing 100081, People's Republic of China}

\date{Submitted ****}

\begin{abstract}
Migratory birds can utilize the geomagnetic field for
orientation and navigation through a widely accepted radical-pair
mechanism. Although many theoretical works have been done, the
available experimental results have not been fully considered,
especially, the temporary disorientation induced by the field which is
increased by $30\%$ of geomagnetic field and the disorientation of
the very weak resonant field of $15nT$. In this paper, we consider
the monotonicity of the singlet yield angular profile as the prerequisite of direction sensitivity,
and find that for some optimal values of the hyperfine coupling parameters, that is the order of $10^{-7}\sim10^{-6}meV$,
the experimental results available by far can be satisfied.
We also investigate the effects of two decoherence environments and demonstrate that,
in order to satisfy the available experimental results,
the decoherence rate should be lower than the recombination rate.
Finally we investigate the effects of the fluctuating magnetic noises, and find
that the vertical noise destroys the monotonicity of the profile completely,
but the parallel noise preserves the monotonicity perfectly and even can enhance the direction sensitivity.
\end{abstract}


\maketitle

\section{Introduction}
Recently, a new interdisciplinary subject called \textit{quantum biology} \cite{quantum biology1,quantum biology2} arouses growing interests in scientists. The major purpose of this subject is to understand the biological phenomena using the fundamental theory of quantum mechanics,
such as photosynthesis \cite{Sarovar,Engel,Collini,Plenio,Mohseni,Briggs,Ringsmuth,Nalbach}, natural selection \cite{Lloyd},
the process of olfaction \cite{Turin,Brookes}, enzymatic reactions \cite{Harkins,Canfield}, and avian magnetoreception \cite{Kominis,Jianming,Jones,Gauger,Gauger2,Caijianming,CYCai,Cai,Jayendra,Hogben,Stoneham}.
Here, we are specifically interested in avian magnetoreception.

It is well known that certain migratory birds can use the Earth's
magnetic field for orientation and navigation
\cite{Mouritsen,Johnsen,Ritz} through a widely accepted radical-pair mechanism
\cite{Steiner,Wiltschko,Lohmann,Maeda,Rodgers} which was first
proposed in the pioneering work by Klaus Schulten \emph{et al}.
\cite{Schulten}. Based on such mechanism, an avian compass model
has been proposed theoretically \cite{Ritz2000}.
The behavioral experiment showed that the avian compass is an inclination compass,
i.e., it is sensitive to the axis but not to the polarity of the geomagnetic field \cite{Wiltschko1972}.
Furthermore, the direction sensitivity is limited to a narrow ``functional window":
the avian compass is disoriented temporarily when the intensity of local magnetic field increases
or decreases by about $30\%$ of the local geomagnetic field,
however, reworks after a sufficiently long time to
adapt itself \cite{Wiltschko1972,Wiltschko1978,Wiltschko2006}.
Compared with the geomagnetic field of $46\mu T$,
the resonance magnetic field of $480nT$ which is not parallel to the geomagnetic field
disrupts the bird's ability to orient \cite{Ritz2004,Thalau}.
In addition, the disorientation induced by the resonance magnetic fields of
$150nT$, $48nT$ and even $15nT$ has also been observed, while
the resonance field of $5nT$ does not disrupt the birds \cite{Ritz2009}.
If the birds are exposed to a stronger local magnetic field of $92\mu T$,
resonance magnetic fields of $150nT$, $48nT$ and $15nT$
still disrupt the birds, and still that of $5nT$ does not \cite{Ritz2009}.

Recently, both the lifetime and the coherence time of the radical pair were discussed.
Considering the fact that the $150nT$ resonance field disrupts the birds, Erik \emph{et al}.
estimated the lifetime and coherence time for a certain hyperfine (HF) coupling that is greater than
the intensity of the geomagnetic field \cite{Gauger,Gauger2}.
Besides considering the resonant field of $150nT$, Jayendra \emph{et al.}
additionally investigated the influences
of the magnetic field reduced by $30\%$ of the geomagnetic
field and the resonance fields of $480nT$ and $48nT$ for a HF
coupling that approximately equals the intensity of geomagnetic
field \cite{Jayendra}. Generally the authors chose a specific value
of the HF coupling parameter and did not give the reason. We have only seen
one theoretical work discussing the effects of HF coupling, and they
optimized the hyperfine coupling parameters to achieve the best magnetic field
sensitivity \cite{Cai}. Although many theoretical works have been
done, until now the experimental results have not been fully considered, for example, the effects of the very weak
oscillating fields of $15nT$ and $5nT$ and that of local magnetic
field which is increased by $30\%$ of the geomagnetic field. Very recently, Erik \emph{et al}.
have pointed out that the effect of the resonant field of $15nT$
should be considered \cite{Gauger2}. In this paper we consider
all the experimental features mentioned above: (\romannumeral1) the fields which are decreased and
increased by about $30\%$ of the geomagnetic field induce the
transient disorientation; (\romannumeral2)
the additional resonant fields of $480nT$, $150nT$, $48nT$ and even $15nT$ which
are orthogonal to the local geomagnetic field cause the
disorientation, but that of $5nT$ does not; (\romannumeral3) if the birds are exposed to a stronger local magnetic field of $92\mu T$,
the resonance magnetic fields of $150nT$, $48nT$ and $15nT$
still disrupt the birds, and the $5nT$ resonance field does not.
Based on these experimental results, we estimate the HF coupling parameters.
We find that the HF coupling parameters should be the order of $10^{-7}\sim10^{-6}meV$
and find the optimal values of the HF coupling parameters so that all the experimental
results mentioned above can be satisfied.
The intriguing feature of the avian compass is that it can work using the fundamental theory of
quantum mechanics at room temperature when various kinds of noises may exist.
Here we also investigate the effects of the two environment noises on the singlet yield.
We demonstrate that, in order to satisfy all the available experimental results,
the decoherence rate should be lower than the recombination rate.
Finally, we also investigate the effects of the random
fluctuation of the magnetic field on the avian compass because it is inevitable around the world, and find
that the monotonicity is destroyed completely by vertical noise but preserved by parallel noise.
Moreover, the parallel noise even can enhance the direction sensitivity of the avian compass.

This paper is organized as follows. In section \uppercase\expandafter{\romannumeral2},
we introduce the most basic avian compass model. Next, we discuss the effects of the HF coupling
on the singlet yield in section \uppercase\expandafter{\romannumeral3}
and the effects of decoherence noises in section \uppercase\expandafter{\romannumeral4}.
Then, we investigate the effects of the random fluctuating
magnetic field in section \uppercase\expandafter{\romannumeral5}
and discuss the recombination rate $k$ in section \uppercase\expandafter{\romannumeral6}.
Finally, some discussions and conclusions are given in section \uppercase\expandafter{\romannumeral7}.
\section{Model}

\label{model}
The most basic model of the avian compass consists of two electronic spins
coupled to an external magnetic field and one nuclear spin.
The nucleus spin interacts anisotropically with only one of the electron spins,
thus it provides asymmetry and leads to singlet-triplet transition required for
the direction sensitivity. The corresponding Hamiltonian is
\begin{equation}\label{H1}
    \hat{H}=\hat{I}\cdot A\cdot\hat{S_{1}}+\gamma \textbf{B}\cdot(\hat{S_{1}}+\hat{S_{2}}),
\end{equation}
where $\hat{I}$ is the nuclear spin operator, and $A$ is the anisotropic hyperfine (HF) tensor
with a diagonal form $A=diag(A_{x},A_{y},A_{z})$.
And we consider an axially symmetric molecule, i.e., $A_{x}=A_{y}$.
$\hat{S_{i}}\equiv(\sigma^{i}_{x},\sigma^{i}_{y},\sigma^{i}_{z})$
are the electronic spin operators ($i=1,2$), $\gamma=\frac{1}{2}\mu_{B}g_{s}$ is the gyromagnetic ratio,
with $\mu_{B}$ is the Bohr magneton and $g_{s}$ is the $g$-factor of the electron.
Here we assume that the $g$-factors are the same for both electron spins and set their values according to
free electron, i.e., $g_{s}=2$. $\textbf{B}$ is the external magnetic field around the radical pair.
We consider a geomagnetic field $\textbf{B}_{\textbf{0}}$ plus
a resonant radio frequency field $\textbf{B}_{\textbf{rf}}$:
\begin{equation}\label{B}
\begin{split}
   \textbf{B}&=\textbf{B}_{\textbf{0}}+\textbf{B}_{\textbf{rf}} \\
   &=B_{0}(\sin\theta \cos\phi, \sin\theta \sin\phi, \cos\theta) \\
   &+B_{rf}\cos\omega t(\sin\alpha \cos\beta, \sin\alpha \sin\beta, \cos\alpha).
\end{split}
\end{equation}
$B_{0}$ is the intensity of the Earth's magnetic field, and $\theta$ and $\phi$ describe its orientation to the basis of the HF tensor. $B_{rf}$ is the strength of additional oscillating field with frequency $\omega$.
$\alpha$ and $\beta$ give the direction of the oscillating field.
Due to the axial symmetry of the HF tensor we set $\phi=0$ and focus on $\theta\in[0,\pi/2]$
without loss of generality. This is supported by the experiment that the avian compass
does not depend on the polarity of magnetic field but only on its inclination \cite{Wiltschko1972}. In this paper we only consider the vertical oscillating field
with $\alpha=\frac{\pi}{2}-\theta$ and $\beta=0$,
because the parallel oscillating field has no effect on avian compass \cite{Ritz2004,Thalau,Ritz2009}.

We consider the same singlet and triplet recombination rates, i.e., $k_{S}=k_{T}=k$,
and in this situation, the singlet yield can be calculated as
\begin{equation}\label{singlet yield}
    \Phi_{s}=\int_{0}^{\infty}r(t)f_{s}(t)dt,
\end{equation}
where $r(t)=k\exp(-kt)$ is the radical recombination probability distribution \cite{Steiner},
and $f_{s}(t)=\langle S|\rho_{s}(t)|S\rangle$ is the population of the singlet state
$|S\rangle=\frac{1}{\sqrt{2}}(|01\rangle-|10\rangle)$.
$\rho_{s}(t)$ is the reduced
electronic spin state at time $t$ with the partial trace over the nucleus subspace.
Recently, it has been pointed out that the lifetime of radical pair
should be the order of $10^{-4} s$, i.e., $k=10^{4} s^{-1}$ \cite{Gauger, Gauger2}.
In this paper we let $k=10^{4} s^{-1}$ and will discuss the validity of it in Sec. \uppercase\expandafter{\romannumeral6}.
We suppose that the electronic spins are initially in the singlet state $|S\rangle$
and the nucleus is in a completely mixed state, i.e.,
$\rho(0)=\frac{1}{2}(|S,\uparrow\rangle\langle S,\uparrow|+|S,\downarrow\rangle\langle S,\downarrow|)$.
\section{estimation of the HF coupling parameters}
Are there any appropriate values of the HF coupling parameters consistent
with all the available experimental results
for this basic model of avian compass?
To answer this question, we consider all the experimental results mentioned above
and investigate the roles of the HF coupling parameters played on the singlet yield angular profile.
First of all, we strictly consider the monotonicity of the singlet yield $(\Phi_{s})$ angular
profile as the prerequisite of direction sensitivity. We argue that if the
singlet yield varies non-monotonously with the direction angle, the
same signal will be induced for different directions,
and the disorientation will occur.

Now without considering the environment we investigate a ``cigar-shaped HF tensor", i.e., $A_{z}>A_{x}=A_{y}$.
The geomagnetic field is set as $B_{0}=46\mu T$ which is the intensity of the geomagnetic field in Frankfurt \cite{Ritz2009}.
For the convenience of our calculation,  we consider $\gamma B_{0}$ as the energy scale with $B_{0}=46\mu T$.
First we let $A_{x}=A_{y}=0$ and investigate the role of the vertical factor $A_{z}$.
Without considering the resonant magnetic field, i.e., $B_{rf}=0$,
the analytic result can be obtained from Eq. \eqref{singlet yield} \cite{Cai}:
$\Phi_{s}(\theta)=\frac{1}{2}[\Phi_{s}(\theta,A_{z})+\Phi_{s}(\theta,-A_{z})]$ with
$\Phi_{s}(\theta,a)=\frac{1}{4}(1+c^2)+\frac{1}{4}(1-c^2)[g(B_1)+g(B_0)]
+\frac{1}{8}(1-c)^{2}g(B_1+B_0)+\frac{1}{8}(1+c)^{2}g(B_1-B_0)$,
where $c=\cos(\theta-\theta^{'})$, $g(x)=k^2/(k^2+x^2)$,
$B_{1}^{2}=(B_0\cos\theta+a)^2+B_{0}^{2}\sin^{2}\theta$,
$\sin\theta^{'}=B_{0}\sin\theta/B_{1}$, and $\cos\theta^{'}=(B_{0}\cos\theta+a)/B_{1}$.
From our numerical calculations we find that $A_{z}$ can be approximately divided into three regimes:
(1) very weak hyperfine coupling regime $A_{z}/\gamma B_{0}\in(0,~2\times10^{-3})$,
in which the singlet yield increases
monotonously with $\theta$;
(2) strong hyperfine coupling regime $A_{z}/\gamma B_{0}>3$,
in which the singlet yield decreases monotonously with the direction angle;
(3) transition regime $A_{z}/\gamma B_{0}\in(2\times10^{-3},~3)$,
in which the singlet yield profile changes from increasing with $\theta$ to decreasing.

Then we investigate the effect of resonant field and that of magnetic fields which are
increased and decreased by $30\%$ of the geomagnetic field, i.e., $32.2\mu T$ and $59.8\mu T$,
for different values of HF coupling parameters.
Generally we believe that if a resonant field disorients
the birds, a stronger resonant field disrupts them as well.
It is known that the weakest resonant field which disorients the birds is
$15nT$ \cite{Ritz2009}, therefore we set $B_{rf}=15nT$.
When $B_{0}=46\mu T$, the corresponding Larmor frequency is about $1.315MHz$, thus we set $\omega/2\pi=1.315MHz$.
\includegraphics{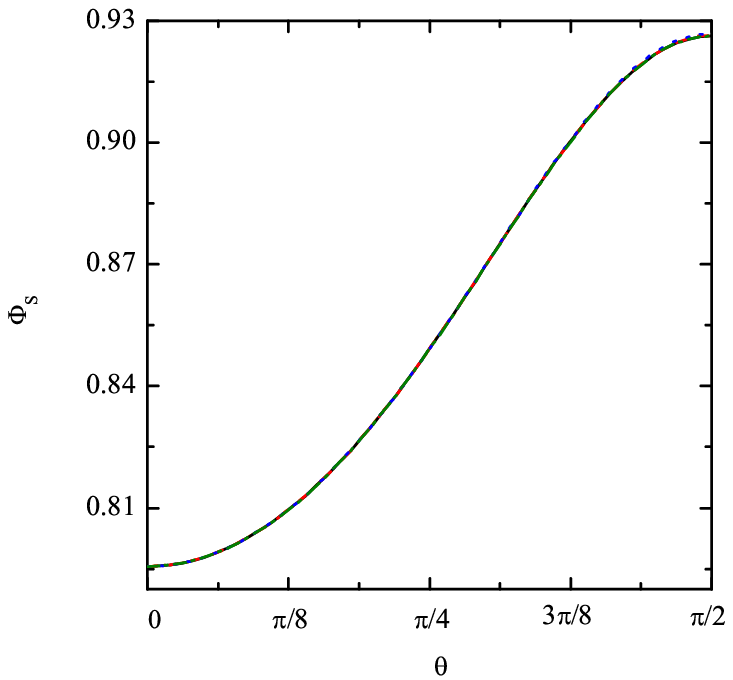}
\begin{center}
\parbox{8cm}{\small{\bf Fig. 1} (Color online) The singlet yield $\Phi_{s}$ as a function of the direction angle $\theta$
for $B_{rf}=5nT$ (red dash) and $15nT$ (blue dot)
compared with the reference value $B_{rf}=0nT$ (black solid).
And the singlet yield $\Phi_{s}$ for $30\%$ stronger (olive dash dot)
and weaker (orange dash dot dot) magnetic field
without considering the resonant field, i.e., $B_{rf}=0nT$.
$B_{0}=46\mu T$, $\omega/2\pi=1.315MHz$ and $A_{z}=2A_{x}=2A_{y}=10^{-3}\gamma B_{0}$.
It is noted that all the singlet yield profiles are coincident with each other.}
\end{center}

In the very weak regime, i.e., $A_{z}/\gamma B_{0}\in(0,~2\times10^{-3})$,
without considering the horizontal factors, i.e., $A_{x}=A_{y}=0$,
the singlet yield increases monotonously with the direction angle,
and the singlet yield profile is regular and stable that
the very weak oscillating field of $15nT$
and $30\%$ weaker and stronger magnetic fields
can not influence it.
Furthermore we consider the values of horizontal factors $A_{x}=A_{y}\neq0$.
According to our numerical calculation we find that for a fixed $A_{z}$,
when $A_{x}=A_{y}<A_{z}$ the singlet yield angular profiles are very similar to that of $A_{x}=A_{y}=0$,
i.e., the singlet yield increases monotonously with the direction angle,
and is immune to the very weak oscillating field of $15nT$
and $30\%$ weaker and stronger magnetic fields.
Here, we set $A_{z}=2A_{x}=2A_{z}=10^{-3}\gamma B_{0}$ as an example and plot the singlet angular profiles in Fig. 1.
From Fig. 1 we can see that the different singlet yield profiles in various cases are almost coincident with each other,
which means that the $15nT$ resonant field and $30\%$ weaker and stronger fields can not disrupt the magnetic sensitivity.
It can be understood that for the very weak HF coupling, the geomagnetic field
plays a dominant role in the dynamics of radical pair,
and the transition rate between singlet and triplet states is very small.
So that the resonance field of $15nT$ and $30\%$ weaker and stronger magnetic fields
can not induce obvious effects.
If $A_{x}~(=A_{y})$ infinitely approaches $A_{z}$ the singlet yield is not angle-dependent any more,
because the anisotropic of hyperfine coupling is destroyed
and the transition between the singlet and triplet states is not allowed.
From the discussion above, the suitable values of horizontal HF factors $A_{x}$ and $A_{y}$ for different $A_{z}$
which are consistent with the experimental results mentioned above
can not be found in this very weak hyperfine coupling regime.

In the strong hyperfine coupling regime of $A_{z}/ \gamma B_{0}>3$, when we do not consider the horizontal
HF coupling factors, i.e., $A_{x}=A_{y}=0$, the singlet yield profile is regular
and decreases monotonously with the direction angle.
The resonant field of $15nT$ can not influence the singlet yield, and
the obvious effects of $30\%$ weaker and stronger fields can not be observed neither.
Furthermore, we consider the values of horizontal factors $A_{x}=A_{y}\neq0$.
For a fixed $A_{z}$ we consider $A_{x}$ (=$A_{y}$) from $0$ to $A_{z}$,
and numerically calculate the singlet yield from Eq. (3).
According to our numerical calculation we find that if the strength of $A_{x}$ ($=A_{y}$) increases to some values,
the obvious influences of $30\%$ weaker and stronger magnetic fields can be observed,
but the effect of the oscillatory field of $15nT$ can still not be observed.
Here, we set $A_{z}=5A_{x}/3=5A_{y}/3=5\gamma B_{0}$ as an example and show the results in Fig. 2.
From Fig. 2 it can be seen that, the resonant fields of $15nT$ and $5nT$
can not influence the singlet yield, however, the influences of $30\%$ weaker
and stronger magnetic fields will be observed.
From our numerical calculations we find that the singlet
yields for the small angles vary slightly with the HF coupling parameter in this strong hyperfine coupling regime.
On the contrary, the singlet yields for large angles, especially for $\theta\approx\frac{\pi}{2}$, vary apparently.
Moreover changing the geomagnetic field is similar to changing the HF coupling parameters,
so that $30\%$ weaker and stronger magnetic field
can influence the singlet yield and induce obvious effects in this strong hyperfine coupling regime.
In this regime, the strong hyperfine coupling plays a dominant role,
and the oscillatory field of $15nT$ is relatively so
weak that it can not induce evident effects.
If we further increase $A_{x}$ ($=A_{y}$), the effect of resonant field of $15nT$ still not be observed, and
the influences of $30\%$ weaker and stronger fields can be observed, but the non-monotonicity will appear.
Certainly if $A_{x}$ ($=A_{y}$) infinitely approaches $A_{z}$ the singlet yield is angle-independent,
because the anisotropic of hyperfine coupling is destroyed
and the transition between the singlet and triplet states is not allowed.
As a result, the appropriate values of horizontal factors $A_{x}$ and $A_{y}$ that agree with all the experimental results
for different $A_{z}$ can not be found in this strong hyperfine coupling regime.
\includegraphics{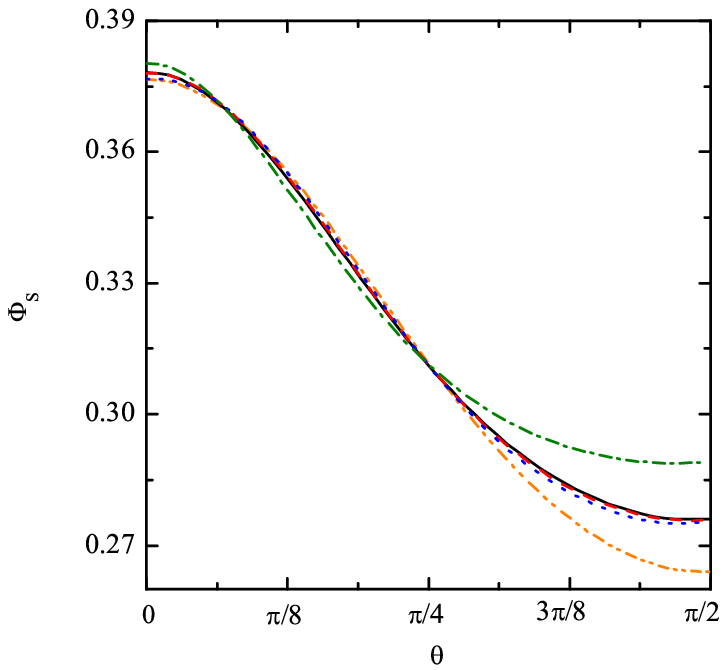}
\begin{center}
\parbox{8cm}{\small{\bf Fig. 2}
(Color online) The singlet yield $\Phi_{s}$ as a function of the direction angle $\theta$
for $B_{rf}=5nT$ (red dash) and $15nT$ (blue dot)
compared with the reference value $B_{rf}=0nT$ (black solid).
And the singlet yield $\Phi_{s}$ for $30\%$ stronger (olive dash dot)
and weaker (orange dash dot dot) magnetic field
without considering the resonant field, i.e., $B_{rf}=0nT$.
$B_{0}=46\mu T$, $\omega/2\pi=1.315MHz$ and $A_{z}=5A_{x}/3=5A_{y}/3=5\gamma B_{0}$.}
\end{center}

In the transition regime of $A_{z}/\gamma B_{0}\in(2\times10^{-3},~3)$,
without considering horizontal factors, i.e., $A_{x}=A_{y}=0$, we calculate $\Phi_{s}$
for different $A_{z}$. Generally, the singlet yield profile changes from increasing with the direction angle to
decreasing. Then we consider the values of $A_{x}=A_{y}\neq0$ for different $A_{z}$ in this regime.
Generally, the singlet yield angular profile is very complex and strongly depends on the values of $A_{x}$, $A_{y}$ and $A_{z}$.
According to different characteristics of the singlet yield angular profile,
the transition regime can be further divided into three sub-regimes:
(a) $A_{z}/\gamma B_{0}\in(2\times10^{-3},~0.1)$;
(b) $A_{z}/\gamma B_{0}\in(0.1,~1)$;
(c) $A_{z}/\gamma B_{0}\in(1,~3)$.

In sub-regime (a), we select more than ten values of $A_{z}$.
For each $A_{z}$, we consider more than ten values of $A_{x}=A_{y}\in(0,~A_{z})$,
and calculate $\Phi_{s}$ from Eq. (3).
It can be concluded from our numerical calculations that we can always find
some appropriate values of $A_{x}$ and $A_{y}$ for any fixed $A_{z}$ in this sub-regime that the monotonicity of
the singlet yield profile can be observed.
Moreover we can find some appropriate values of $A_{x}$ and $A_{y}$ for different $A_{z}$ so that
the influence of the resonance field of $15nT$ can be observed. However, the effects of
$30\%$ weaker and stronger fields can not be found in this sub-regime.
In sub-regime (b), similar to sub-regime (a), we numerically calculate the singlet yield from Eq. (3).
From numerical calculations we can always find appropriate
values of $A_{x}$ and $A_{y}$ for any fixed $A_{z}$ so that the fields which are decreased and
increased by about $30\%$ of the geomagnetic field can induce the
transient disorientation and the $15nT$
orthogonal oscillating field can disrupt the birds.
It is known that the resonant field of $5nT$ does not disturb the orientation \cite{Ritz2009}.
From our numerical calculations we also find that for the above appropriate
values of $A_{x}$ and $A_{y}$ with the fixed $A_{z}$ the $5nT$ resonant field changes the angular profile so slightly
that the disorientation can not be induced.
\includegraphics{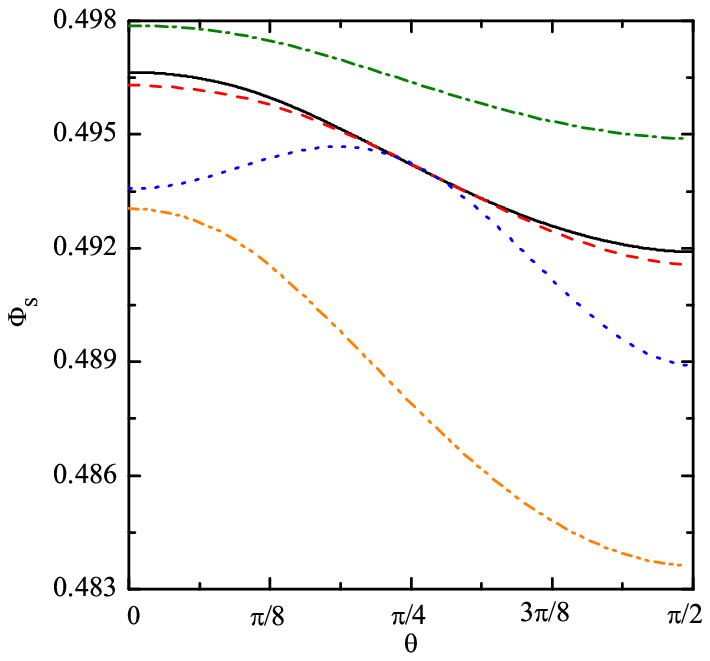}
\begin{center}
\parbox{8cm}{\small{\bf Fig. 3}
(Color online) The singlet yield $\Phi_{s}$ as a function of the direction angle $\theta$
for $B_{rf}=5nT$ (red dash) and $15nT$ (blue dot)
compared with the reference value $B_{rf}=0nT$ (black solid).
And the singlet yield $\Phi_{s}$ for $30\%$ stronger (olive dash dot)
and weaker (orange dash dot dot) magnetic field
without considering the resonant field, i.e., $B_{rf}=0nT$.
$B_{0}=46\mu T$, $\omega/2\pi=1.315MHz$ and $A_{z}=2A_{x}=2A_{y}=\gamma B_{0}/6$.}
\end{center}
Here we take $A_{z}=2A_{x}=2A_{y}=\gamma B_{0}/6$ as
an example, and show the results in Fig. 3. It can be seen from Fig. 3 that
$30\%$ weaker and stronger magnetic fields change the singlet yield significantly,
but the monotonicity preserves perfectly, so that they induce disorientation transiently
and the avian compasses could rework after a sufficiently long time to adapt themselves.
The resonant field of $15nT$ destroys the monotonicity so that the magnetic sensitivity will be disrupted,
and in contrast, the resonant field of $5nT$ changes the singlet yield so slightly that the disorientation can not be induced.
Experimentally, the resonant fields of $480nT$, $150nT$ and $48nT$ also disrupt the birds \cite{Ritz2004,Thalau,Ritz2009}.
Thus we also consider the resonant fields of $480nT$, $150nT$ and $48nT$.
Consistently, all the profile monotonicity is destroyed, i.e., the magnetic sensitivity is disrupted.
In sub-regime (c), similar to the calculations of sub-regime (a),
we can find the appropriate values of $A_{x}$ and $A_{y}$ that the obvious influences of $30\%$ weaker and stronger field
can be observed, but the singlet yield angular profiles are always non-monotonous.
And the oscillating field of $15nT$ can not induce significant effects in this sub-regime.
So far we have considered all the values of the hyperfine coupling parameters and for clarity the results are summarized in TABLE \ref{table:results}.
Without considering the environment, from the discussion above we can conclude that the appropriate values of hyperfine coupling parameters
which are consistent with all the available experimental results can always
be found in sub-regime (b) $A_{z}/\gamma B_{0}\in(0.1,~1)$,
and the corresponding hyperfine factor $A_{z}\in (2.66\times10^{-7}meV, \ 2.66\times10^{-6}meV)$.
Moreover from numerical calculations we find that generally the appropriate values of $A_{x}$ and $A_{y}$ increase with $A_{z}$.
If we consider the mechanism of the interaction between the electron and nucleus spins,
and let $a=\mu_{0}\mu_{B}\mu_{N}/(4\pi a^{3}_{0})$ as the energy scale,
where $\mu_{B}$ and $\mu_{N}$ are the Bohr magneton and nuclear magneton respectively,
$\mu_{0}$ is the electric permittivity of free space, and $a_{0}$ is the Bohr radius for hydrogen.
In this case, the optimal values of $A_{z}/a\in(1.35\times10^{-3}, \ 1.35\times10^{-2})$.
\begin{widetext}
\begin{center}
\begin{table}
\caption{The results for different hyperfine coupling regimes.
Y: The appropriate values of the parameters can be found to exhibit the
corresponding characteristics; N: the appropriate values of the parameters can not be found to exhibit the
corresponding characteristics.} 
\begin{tabular}{|c|c|c|c|}\hline
  regimes ($A_{z}/ \gamma B_{0}\in$)         & monotonicity & effect of $15nT$ resonant field              & effects of $30\%$ weaker and stronger fields \\ \hline
  $(0, 2\times10^{-3})$                      & Y            & N                                  & N                                 \\ \hline
  $(2\times10^{-3}, 0.1)$                    & Y            & Y                                  & N                                 \\ \hline
  $(0.1, 1)$                                 & Y            & Y                                  & Y                                 \\ \hline
  $(1, 3)$                                   & N            & N                                  & Y                                 \\ \hline
  $>3$                                       & Y            & N                                  & Y                                 \\ \hline
\end{tabular}
\label{table:results}
\end{table}
\end{center}
\end{widetext}

\includegraphics{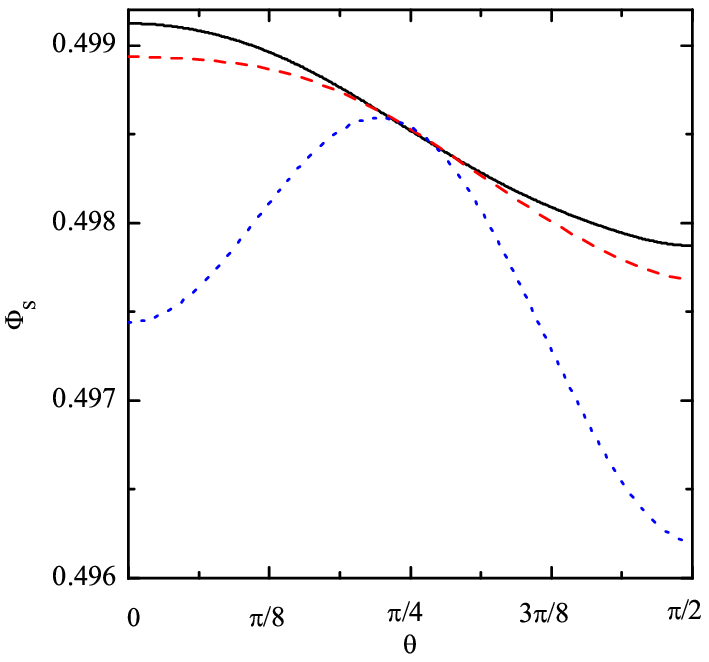}
\begin{center}
\parbox{8cm}{\small{\bf Fig. 4}
(Color online) The singlet yield $\Phi_{s}$ as a function of the direction angle $\theta$
for $B_{rf}=5nT$ (red dash) and $15nT$ (blue dot)
compared with the reference value $B_{rf}=0nT$ (black solid).
$B^{'}_{0}=92\mu T$, $\omega/2\pi=2.63MHz$ and $A_{z}=2A_{x}=2A_{y}=\gamma B_{0}/6$ with $B_{0}=46\mu T$.}
\end{center}

Moreover as the behavioral experiments shown, if the birds are exposed to a stronger local field of
$B^{'}_{0}=92\mu T$ whose corresponding Larmor frequency is $2.63MHz$,
the resonance field of $15nT$ still disorients the birds, but
that of $5nT$ does not \cite{Ritz2009}.
Accordingly, we replace the geomagnetic field by a stronger field of $92\mu T$ and consider the same HF coupling parameters above,
i.e., $A_{z}=2A_{x}=2A_{y}=\gamma B_{0}/6$. For the resonant fields of $15nT$ and $5nT$,
we calculate $\Phi_{s}$ and show the results in Fig. 4.
It can be seen from Fig. 4 that when the resonant field is $15nT$, the
monotonicity of the singlet yield is destroyed, and the magnetic sensitivity is disrupted.
However, for the resonant field of $5nT$, the change of the singlet yield is so little that the orientation can not be disturbed.

\includegraphics{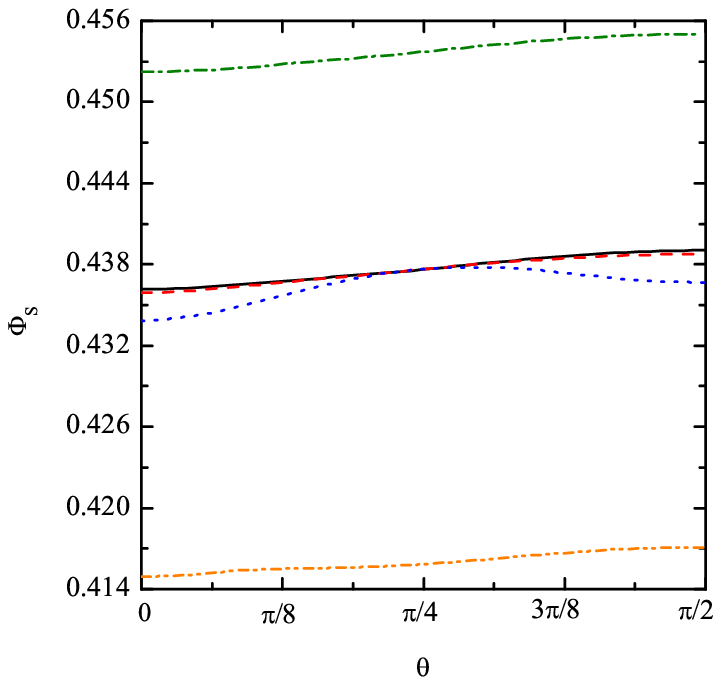}
\begin{center}
\parbox{8cm}{\small{\bf Fig. 5}
(Color online) The singlet yield $\Phi_{s}$ as a function of the direction angle $\theta$
for $B_{rf}=5nT$ (red dash) and $15nT$ (blue dot)
compared with the reference value $B_{rf}=0nT$ (black solid).
And the singlet yield $\Phi_{s}$ for $30\%$ stronger (olive dash dot)
and weaker (orange dash dot dot) magnetic field
without considering the resonant field, i.e., $B_{rf}=0nT$.
$B_{0}=46\mu T$, $\omega/2\pi=1.315MHz$ and $A_{x}=A_{y}=\frac{10}{9}Az=\gamma B_{0}/2$.}
\end{center}

Hitherto, we have investigated the effects of the ``cigar-shaped HF tensor". One may ask whether for the ``disc-shaped HF tensor",
i.e., $A_{x}=A_{y}>A_{z}$, there are appropriate values of the hyperfine parameters consistent with the experimental results mentioned above?
Similar to ``cigar-shaped HF tensor", when the horizontal factor $A_{x}$ ($=A_{y}$) is very weak
the resonant field of $15nT$ and $30\%$ weaker and stronger fields can not influence the magnetic sensitivity.
If the horizontal factor $A_{x}$ ($=A_{y}$) is strong, $30\%$ weaker and stronger fields will influence the singlet yield,
but the oscillating field of $15nT$ will not.
In the intermediate regime, we choose different values of horizontal factor $A_{x}$ ($=A_{y}$).
And for each values of $A_{x}$ ($=A_{y}$), we consider $A_{z}$ from $0$ to $A_{x}$ (=$A_{y}$),
and calculate the singlet yield from Eq. (3).
We find that without considering the environment the appropriate values of $A_{z}$ for different $A_{x}$ ($=A_{y}$),
which are consistent with the available experimental results,
always exist in the regime of $A_{x}/\gamma B_{0}$ ($=A_{y}/\gamma B_{0}$) $\in(0.2,\ 0.7)$.
And the corresponding hyperfine coupling parameter $A_{x}$ ($= A_{y}$) $\in(5.32\times10^{-7}meV,\ 1.862\times10^{-6}meV)$.
In the case of energy scale $a$, the hyperfine parameter $A_{x}/a$ ($=A_{y}/a$) $\in(2.70\times10^{-3},~9.44\times10^{-3})$.
Here we take $A_{x}=A_{y}=\frac{10}{9}Az=\gamma B_{0}/2$ as an example and numerically calculate the singlet yield from Eq. (3).
The results are shown in Fig. 5. It can be seen from Fig. 5 that,
without considering the resonance field, i.e., $B_{rf}=0nT$, the singlet yield increases monotonously with the direction angle.
Under the influence of $15nT$ oscillating field, the monotonicity is destroyed, and the orientation is disrupted.
However, for the resonant field of $5nT$ the singlet yield changes so slightly that the disorientation can not be induced.
$30\%$ weaker and stronger magnetic fields change the singlet yield significantly,
but the monotonicity preserves, so that avian compass will disorient transiently and
rework after a sufficiently long time to adapt itself.
From the discussion above it is known that for both the ``cigar-shaped HF tensor"
and ``disc-shaped HF tensor" without considering the environment
the values of HF coupling parameters should be the order of $10^{-7}\sim10^{-6}meV$.
\section{effects of decoherence}

\begin{widetext}
\begin{center}
\includegraphics{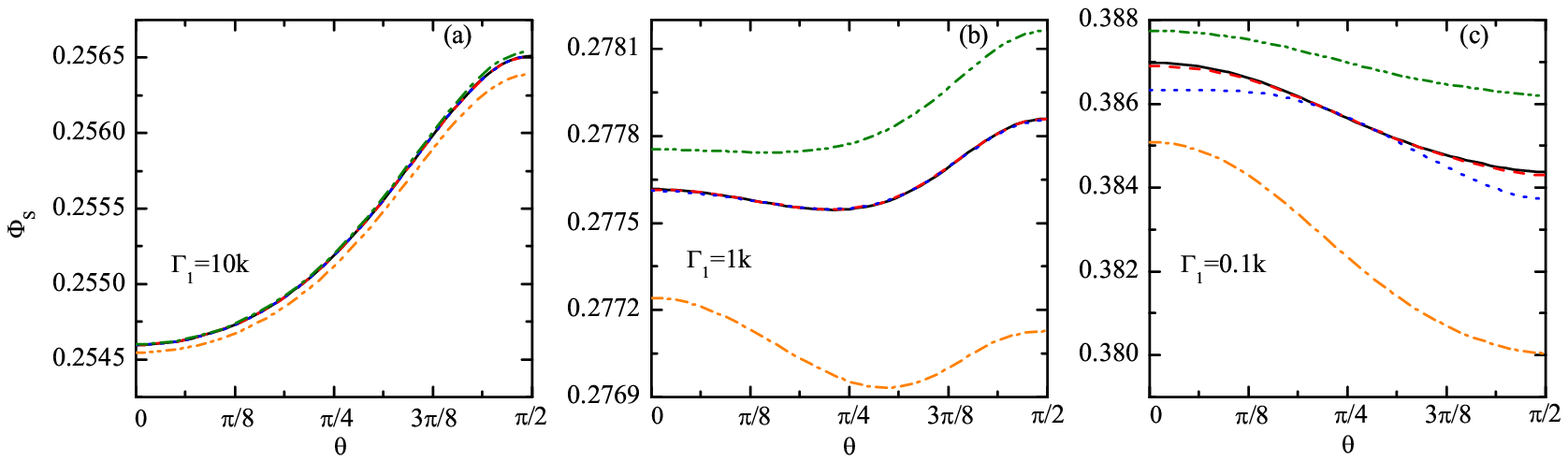}
\parbox{17.2cm}{\small{\bf Fig. 6}
(Color online) The singlet yield $\Phi_{s}$ as a function of the direction angle $\theta$ with different decoherence rate $\Gamma_{1}$
for $B_{rf}=5nT$ (red dash) and $15nT$ (blue dot)
compared with the reference value $B_{rf}=0nT$ (black solid).
And the singlet yield $\Phi_{s}$ for $30\%$ stronger (olive dash dot)
and weaker (orange dash dot dot) magnetic field
without considering the resonant field, i.e., $B_{rf}=0nT$.
$B_{0}=46\mu T$, $\omega/2\pi=1.315MHz$ and $A_{z}=2A_{x}=2A_{y}=\gamma B_{0}/6$.}
\end{center}
\end{widetext}

Decoherence is unavoidable at room temperature. Recently several interesting works have
investigated the effects of decoherence noise \cite{Jianming, Gauger, Cai, Dellis, Kavokin, Tiersch, Walters}.
Firstly we consider the generic noise model \cite{Gauger}:
\begin{equation}\label{dephasing noise}
  \mathcal{L}_{1}(\rho)=\Gamma_{1}\sum\limits_{i} {(L_{i}\rho L^{\dag}_{i}-\frac{1}{2}L^{\dag}_{i}L_{i}\rho-\frac{1}{2}\rho L^{\dag}_{i}L_{i})}
\end{equation}
with the decoherence rate $\Gamma_{1}$,
where the noise operators $L_{i}$ are $\sigma_{x}$, $\sigma_{y}$, $\sigma_{z}$
for each electron spin individually.
We consider the optimal values of HF coupling parameters $A_{z}=2A_{x}=2A_{y}=\gamma B_{0}/6$,
and numerically calculate the singlet yield with different decoherence rates.
Here we take $\Gamma_{1}=10k$, $1k$ and $0.1k$ as examples
and plot their corresponding singlet yield angular profiles in Fig. 6.
From our numerical calculation we find that when $\Gamma_{1}$ is approximately equal to or lager than $10k$
the resonance fields of $15nT$ and $5nT$ and $30\%$ weaker and stronger fields can not influence the
magnetic sensitivity, which can be seen in Fig. 6 (a). If $\Gamma_{1}$ approximately equals $1k$, it can be seen from Fig. 6 (b) that,
$30\%$ weaker and stronger fields will influence the singlet yield significantly,
but the non-monotonicity will arise. And the resonant fields of $15nT$ and $5nT$ can not influence the singlet yield.
If $\Gamma_{1}$ is approximately equal to or lower than $0.1k$, it can be seen from Fig. 6 (c) that
the singlet yield decreases monotonously with the direction angle.
Interestingly, when $\Gamma_{1}\approx0.1k$ the angular profile becomes flat at the small angle regime under the influence of
the resonant field of $15nT$, and if $\Gamma_{1}<0.1k$ the monotonicity will be destroyed.
This means that the $15nT$ oscillating field can disrupt the magnetic sensitivity, but that of $5nT$ can not.
Moreover, $30\%$ weaker and stronger fields influence the singlet yield significantly but preserve the monotonicity perfectly.
In conclusion for the generic noise model, in order to satisfy the available experimental results,
the decoherence rate should be approximately equal to or lower than $0.1k$.

\begin{widetext}
\includegraphics{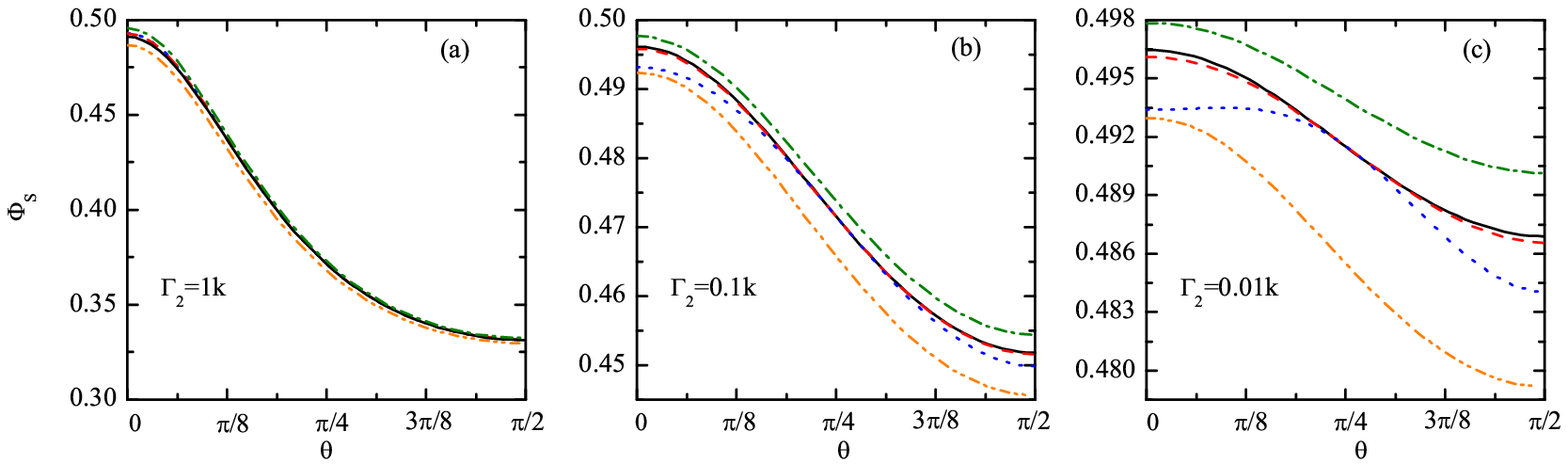}
\begin{center}
\parbox{17.2cm}{\small{\bf Fig. 7}
(Color online) The singlet yield $\Phi_{s}$ as a function of the direction angle $\theta$ with different $\Gamma_{2}$
for $B_{rf}=5nT$ (red dash) and $15nT$ (blue dot)
compared with the reference value $B_{rf}=0nT$ (black solid).
And the singlet yield $\Phi_{s}$ for $30\%$ stronger (olive dash dot)
and weaker (orange dash dot dot) magnetic field
without considering the resonant field, i.e., $B_{rf}=0nT$.
$B_{0}=46\mu T$, $\omega/2\pi=1.315MHz$, $A_{z}=2A_{x}=2A_{y}=\gamma B_{0}/6$, $d=1$.}
\end{center}
\end{widetext}

Then we investigate the correlated and uncorrelated dephasing noises \cite{Cai},
\begin{equation}\label{dephasing noise}
  \mathcal{L}_{2}(\rho)=\frac{1}{4}\sum\limits_{i=1,2} {(2L_{i}\rho L^{\dag}_{i}-L^{\dag}_{i}L_{i}\rho-\rho L^{\dag}_{i}L_{i})}.
\end{equation}
The noise operators are $L_{1}=(\frac{\Gamma_{2}}{1+d^{2}})^{1/2}[\sigma^{(1)}_{z}+d\sigma^{(2)}_{z}]$ and $L_{2}=(\frac{\Gamma_{2}}{1+d^{2}})^{1/2}[d\sigma^{(1)}_{z}+\sigma^{(2)}_{z}]$,
where $\sigma_{z}$ is Pauli operator. The parameter $d$
characterizes how correlated is the dephasing, i.e., $d=0$
for uncorrelated dephasing and $d=1$ for a perfectly correlated one.
We still set $A_{z}=2A_{x}=2A_{y}=\gamma B_{0}/6$ as an example and numerically calculate the singlet yield
under these noises with different decoherence rates. From our calculations we find that
the correlated and uncorrelated dephasing noises have similar influences on the singlet yield
when decoherence rate changes from $0.01k$ to $1k$. Therefore we only show the effects of the correlated noise.
Here, we take $\Gamma_{2}=1k$, $0.1k$ and $0.01k$
as examples and plot their corresponding singlet angular profiles in Fig. 7.
From our numerical calculation we find that when $\Gamma_{2}$ is approximately equal to or lager than $1k$,
the resonant field of $15nT$ and $30\%$ weaker and stronger fields can not influence the singlet yield,
which can be seen in Fig. 7 (a).
If $\Gamma_{2}$ approximately equals $0.1k$ it can be seen from Fig. 7 (b) that,
$30\%$ weaker and stronger fields do influence the singlet yield but not greatly.
And the oscillating fields of $15nT$ can not influence the singlet yield.
When $\Gamma_{2}$ is approximately equal to or lower than $0.01k$, it can be seen from Fig. 7 (c) that
$30\%$ weaker and stronger magnetic fields change the singlet yield significantly, but preserve the monotonicity perfectly,
so that avian compass disorients transiently and reworks after a sufficiently long time to adapt itself.
The resonant field of $15nT$ destroys the monotonicity so that the magnetic sensitivity is disrupted,
and in contrast, the resonant field of $5nT$ changes the singlet yield so slightly that the disorientation can not be induced.
Thus in order to satisfy the available experimental results,
the decoherence rate of the dephasing noise model should be approximately equal to or lower than $0.01k$.
We can conclude that the values of HF coupling parameters should be the order of $10^{-7}\sim10^{-6}meV$ as long as the decoherence rate is approximately equal to or lower than $0.1k$ for the generic noise model, and is approximately equal to or lower than $0.01k$ for the correlated and uncorrelated dephasing noise model.

In this paper, we assume the monotonicity of the singlet yield angular profile as the prerequisite of direction sensitivity,
adopt the basic model of the avian compass, and consider the common noise models.
And we find that in order to satisfy all the experimental results mentioned above,
especially the effect of 15$nT$ oscillating field,
the decoherence rate should be the order of $0.1k$ for the generic noise model and $0.01k$ for the dephasing noise model respectively.
The low decoherence rate of those noise models might be understood as follows.
(1) As we have mentioned above the lifetime of radical pair
should be the order of $10^{-4} s$, i.e., $k=10^{4} s^{-1}$ \cite{Gauger, Gauger2},
and we will further give the reason why $k$ should be the order of $10^{4}s^{-1}$ in section VI.
We believe that the decoherence rate generally should be less than the lifetime of radical pair,
so the low dcoherence rate is plausible. (2) The experiment showed that the resonant field of $15nT$ disrupts the birds,
which is extremely weak relative to the geomagnetic field of $46\mu T$.
It was pointed out that the slow spin flip time which is induced
by the resonant field of $15nT$ implies that the process it disrupts must be slow too \cite{Walters},
so the coherence time should be the order of milliseconds. What kind of noise model is appropriate for the avian compass model really needs further study both theoretically and experimentally, and in this way the mechanism of the long coherence time in the radical pair might be understood completely. Actually the electron spin relaxation time of the molecules has been investigated \cite{Morton1,Morton2,Morton3},
and the coherence time $0.25 ms$ for a molecular electron spin has been reported \cite{Morton1}.
The related problem of decoherence in a singlet/triplet quantum dot has also been studied in Refs. \cite{Johnson,Petta}.

\section{effects of the fluctuating fields}
\label{Fluctuation field}
Besides the intrinsic decoherence noises, there are ubiquitous external magnetic noise around the avian compass.
So we investigate the effect of the fluctuating magnetic
noise on the avian compass. We replace the resonant field by a fluctuating
magnetic field
\begin{equation}\label{FB}
     \textbf{B}^{'}=B^{'}(t)(\sin\vartheta \cos\varphi, \sin\vartheta \sin\varphi, \cos\vartheta),
\end{equation}
where $B^{'}(t)$ describes the strength of the fluctuating field,
$\vartheta$ and $\varphi$ are its direction angles.
We also set $\varphi=0$ due to the axial symmetry of the HF tensor.
Here, two kinds of fields are investigated: the fluctuating fields parallel and vertical to the geomagnetic field.
For the parallel case, $\vartheta=\theta$; and $\vartheta=\frac{\pi}{2}-\theta$ for the vertical case.
The total Hamiltonian can be written as
\begin{equation}\label{H2}
\begin{split}
H &=\hat{I}\cdot A\cdot\hat{S_{1}}+\gamma \textbf{B}_{\textbf{0}}\cdot(\hat{S_{1}}+\hat{S_{2}})+\gamma\textbf{B}^{'}\cdot(\hat{S_{1}}+\hat{S_{2}}) \\
&=\hat{I}\cdot A\cdot\hat{S_{1}}+\gamma \textbf{B}_{\textbf{0}}\cdot(\hat{S_{1}}+\hat{S_{2}})+\gamma B^{'}(t)M(\vartheta) \\
&=H_{0}+H^{'}(t), \\
\end{split}
\end{equation}
with $H_{0}=\hat{I}\cdot A\cdot\hat{S_{1}}+\gamma \textbf{B}_{\textbf{0}}\cdot(\hat{S_{1}}+\hat{S_{2}})$ and
$H^{'}(t)=\gamma B^{'}(t)M(\vartheta)$.
$M(\vartheta)=\sum_{i}\hat{S_{i}}(\vartheta)$, with $\hat{S_{i}}(\vartheta)=sin\vartheta\sigma^{i}_{x}+cos\vartheta\sigma^{i}_{z}$ ($i=1,2$).
In the interaction picture,
the Liouville's equation can be written as ($\hbar=1$)
\begin{equation}\label{Liouville's equation}
\frac{d}{dt}\rho_{I}(t)=-i[H_{I}(t),~\rho_{I}(t)],
\end{equation}
where, $\rho_{I}(t)=e^{iH_{0}t}\rho(t)e^{-iH_{0}t}$ and $H_{I}(t)=e^{iH_{0}t}H^{'}(t)e^{-iH_{0}t}=
\gamma B^{'}(t)M_{I}(\vartheta, t)$ with $M_{I}(\vartheta, t)=e^{iH_{0}t}M(\vartheta)e^{-iH_{0}t}$.
\begin{widetext}
Generally, Eq. (\ref{Liouville's equation}) can be solved by iteration \cite{Loreti,Wang},
\begin{equation}\label{rhot1}
\begin{split}
\rho_{I}(t) &=\rho_{I}(0)-i\int_{0}^{t}dt_{1}\gamma B^{'}(t_{1})[M_{I}(\vartheta, t_{1}),~\rho_{I}(0)] \\
&-\int_{0}^{t}dt_{1}\int_{0}^{t1}dt_{2}\gamma^{2}B^{'}(t_{1})B^{'}(t_{2})[M_{I}(\vartheta, t_{1}),~[M_{I}(\vartheta, t_{2}),~\rho_{I}(0)]]+\cdot\cdot\cdot.
\end{split}
\end{equation}
Due to the random magnetic field, the average density matrix satisfies the following equation:
\begin{equation}\label{rhot2}
\begin{split}
\langle\rho_{I}(t)\rangle &=\rho_{I}(0)-i\int_{0}^{t}dt_{1}\gamma \langle B^{'}(t_{1})\rangle[M_{I}(\vartheta, t_{1}),~\rho_{I}(0)] \\
&-\int_{0}^{t}dt_{1}\int_{0}^{t1}dt_{2}\gamma^{2}\langle B^{'}(t_{1})B^{'}(t_{2})\rangle[M_{I}(\vartheta, t_{1}),~[M_{I}(\vartheta, t_{2}),~\rho_{I}(0)]]+\cdot\cdot\cdot.
\end{split}
\end{equation}
We consider a Gaussian white noise, i.e., $\langle B^{'}(t)\rangle=0$, thus the $n'$th-order correlation
can be written as
\begin{equation}
\langle B^{'}(t_{1})B^{'}(t_{2})\cdot\cdot\cdot B^{'}(t_{n})\rangle=
\begin{cases}
0& \text{if $n$ is odd},\\
\sum\limits_{\begin{subarray}{l}
all~(n-1)!!\\ ~pairings
\end{subarray}}\langle B^{'}(t_{1})B^{'}(t_{2})\rangle\langle B^{'}(t_{3})B^{'}(t_{4})\rangle\cdot\cdot\cdot\langle B^{'}(t_{n-1})B^{'}(t_{n})\rangle& \text{if $n$ is even},
\end{cases}
\end{equation}
with $(n-1)!!=(n-1)(n-3)\cdot\cdot\cdot5\cdot3\cdot1$ \cite{Wolf}.
We assume that $\langle B^{'}(t)B^{'}(\tau)\rangle=\Gamma\delta(t-\tau)$, i.e., the Markovian process, and obtain
\begin{equation}\label{rhot3}
\begin{split}
     \langle\rho_{I}(t)\rangle &=\rho_{I}(0)
    -\int_{0}^{t}dt_{1}\gamma^{2}\Gamma[M_{I}(\vartheta, t_{1}),~[M_{I}(\vartheta, t_{1}),~\rho_{I}(0)]] \\
    &+\int_{0}^{t}dt_{1}\gamma^{4}\Gamma^{2}[M_{I}(\vartheta, t_{1}),~[M_{I}(\vartheta, t_{1}),\int_{0}^{t_{1}}dt_{2}[M_{I}(\vartheta, t_{2}),~[M_{I}(\vartheta, t_{2}),~\rho_{I}(0)]]]]+\cdot\cdot\cdot, \\
\end{split}
\end{equation}
\end{widetext}
which is just the iterative expression of the following differential equation \cite{Loreti,Wang},
\begin{equation}
   \frac{d}{dt}\langle\rho_{I}(t)\rangle=-\gamma^{2}\Gamma[M_{I}(\vartheta, t),~[M_{I}(\vartheta, t),~\langle\rho_{I}(t)\rangle]].
\end{equation}
In the Schr\"{o}dinger picture, it can be written as
\begin{equation}
   \frac{d}{dt}\langle\rho(t)\rangle=-i[H_{0},~\langle\rho(t)\rangle]-\gamma^{2}\Gamma[M(\vartheta),~[M(\vartheta),~\langle\rho(t)\rangle]].
\end{equation}

\includegraphics{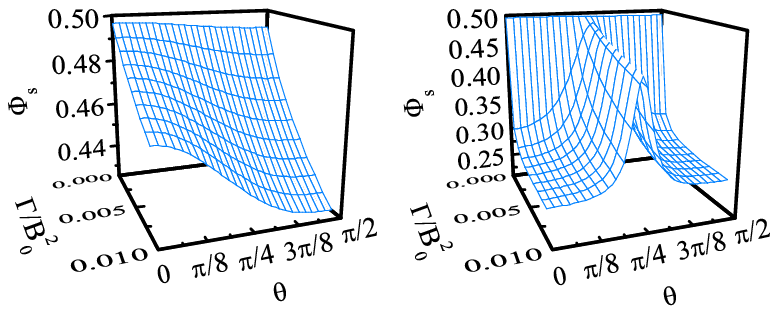}
\begin{center}
\parbox{8cm}{\small{\bf Fig. 8} (Color online) The singlet yield $\Phi_{s}$
as a function of $\Gamma/B_{0}^{2}$ and $\theta$ for the parallel (left) and vertical (right) fluctuating fields respectively.
 $B_{0}=46\mu T$, $A_{z}=2A_{x}=2A_{y}=\gamma B_{0}/6$.}
\end{center}

For $B_{0}=46\mu T$, we numerically calculate the singlet yields
when the fluctuating fields are parallel and
vertical to the geomagnetic field, and
the results are shown in Fig. 8.
It can be seen from Fig. 8 that the vertical fluctuating field destroys the monotonicity of the
singlet yield profile, and thus the avian compass can not work.
In contrast, when the avian compass is exposed to the parallel fluctuating field,
the monotonicity of the singlet yield profile is preserved perfectly.
These results are similar to the results of the parallel and vertical resonant fields.
Furthermore, we can find from Fig. 8 that for the parallel magnetic noise,
all the singlet yields for different angles decrease with the increasing noise
but the difference between the maximum and the minimum singlet yields increases.
Similar to the effects of $30\%$ weaker and stronger magnetic fields,
the significant change of singlet yield might disorient avian compass transiently and
the preservation of the monotonicity will re-orient the avian compass after a sufficiently long time to adapt itself.
Moreover, the increasing of the difference between the maximum and the minimum singlet yields means that
the parallel noise can enhance the direction sensitivity of the avian compass.
\section{Recombination rate of the radical pair}

\begin{widetext}
\includegraphics{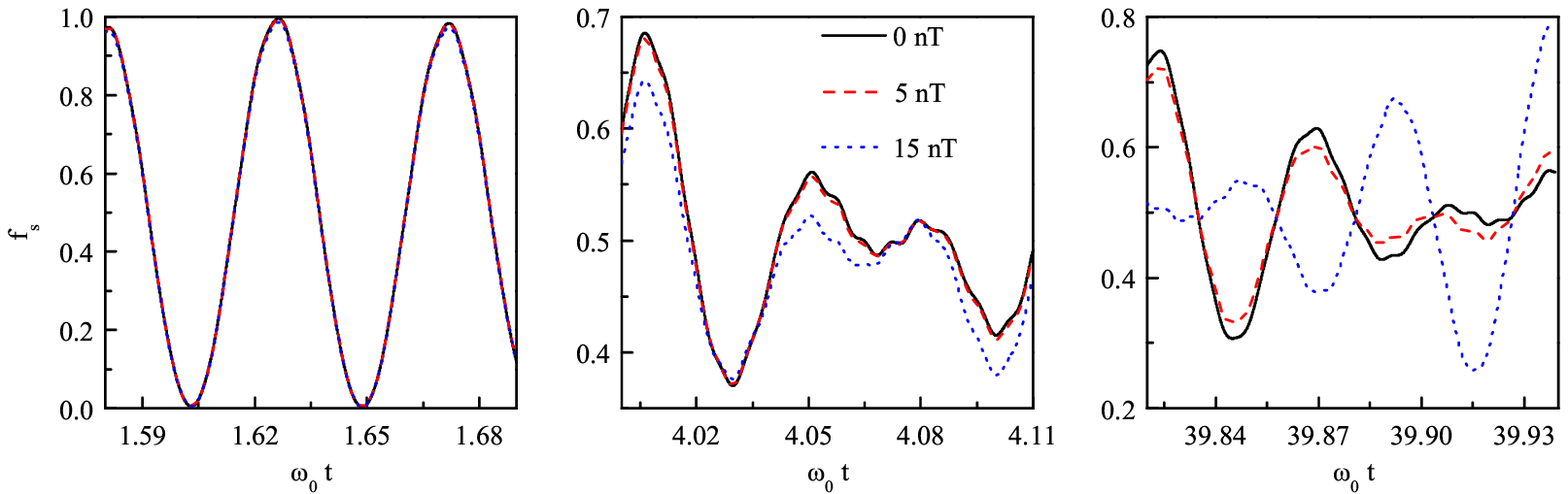}
\begin{center}
\parbox{17.2cm}{\small{\bf Fig. 9} (Color online) The different periods of the evolutions of the singlet state population $f_{s}$
when $B_{rf}=0nT$ (black solid), $5nT$ (red dash) and $15nT$ (blue dot).
$B_{0}=46\mu T$, $\omega/2\pi=1.315MHz$, $\theta=0.1$ and $A_{z}=2A_{x}=2A_{y}=\gamma B_{0}/6$.}
\end{center}
\end{widetext}

Recently, it has been pointed out that the lifetime of radical pair should be the order of $100\mu s$,
i.e., $k=10^{4}s^{-1}$ \cite{Gauger, Gauger2}.
Now, we further discuss the validity of it.
As we know the radical recombination probability distribution $r(t)=k\exp(-kt)$ decays to zero when $t\approx10k^{-1}$,
and there is no singlet yield to be generated after that.
Without considering any decoherence noise and magnetic noise,
we numerically calculate the singlet state population $f_{s}$
under the influences of the additional weak radio frequency fields of $15nT$ and $5nT$.
We let $\omega_{0}=\frac{1}{100\mu s}=10^{4}s^{-1}$  for convenience.
Fig. 9 shows the different periods of evolutions of the singlet state population $f_{s}$ for $B_{rf}=0nT$, $5nT$ and $15nT$
with $B_{0}=46\mu T$, $A_{z}=2A_{x}=2A_{y}=\gamma B_{0}/6$ and $\theta=0.1$.
From Fig. 9, we can see that there are no evident effects of the oscillating fields of $15nT$ and $5nT$ before $t\approx2.0\omega_{0}^{-1}$.
While $r(t)$ decays to zero when $t\approx\omega_{0}^{-1}$ for $k=10^{5}s^{-1}$ and $t\approx0.1\omega_{0}^{-1}$ for $k=10^{6}s^{-1}$.
Thus for both $k=10^{5}s^{-1}$ and $10^{6}s^{-1}$, the resonant fields of $15nT$ and $5nT$ have
no enough time to influence the singlet yield.
The obvious influence of the $15nT$ orthogonal field can be observed after $t\approx3.0\omega_{0}^{-1}$ and
that of the $5nT$ orthogonal field emerges after $t\approx20.0\omega_{0}^{-1}$.
For $k=10^{4}s^{-1}$, $r(t)$
decays to zero when $t\approx10\omega_{0}^{-1}$, therefore the field of $15nT$ can induce obvious effect but the field of $5nT$ can not.
If we consider the case of $k=10^{3}s^{-1}$, $r(t)$ would survive until about $t\approx100\omega_{0}^{-1}$,
as a result, both the influences of the radio frequency fields of $5nT$ and $15nT$ can be observed, i.e.,
the disorientation can be induced by the resonant fields of $5nT$ and $15nT$.
However, as reported by Ritz \emph{et al.}, the most weak intensity of the resonant radio frequency field which disorients the birds is
$15nT$, and the birds will not be disturbed when it is exposed to $5nT$ orthogonal oscillating field \cite{Ritz2009}.
Here, we also consider other direction angles and other values of the HF tensors, and find that
the influence of resonant field of $15nT$ always appears after $t\approx3.0\omega_{0}^{-1}$ and that of the resonant field
of $5nT$ always appears after $t\approx20.0\omega_{0}^{-1}$.
Our investigations clearly show the reason why $k$ should be the order of $10^{4}s^{-1}$.
\section{Conclusions}
Although many theoretical works on avian compass have been
done, until now the experimental results have not been fully considered, and according to our knowledge the effects of the HF
coupling parameters have not been fully considered. In this paper
based on the available experimental results by far, we have estimated the values of the HF coupling parameters.
We have found the optimal values of the HF coupling parameters, which should be the order of $10^{-7}\sim10^{-6}meV$,
so that all the available experimental results can be satisfied.
Furthermore, we also investigate different decoherence models and demonstrate that,
in order to satisfy all the available experimental results by far
for the general noise model the decoherence rate should be equal to or less than $0.1k$,
while for dephasing noise model the decoherence rate should be equal to or less than of $0.01k$.
Due to the inevitable random magnetic noise around the world,
we have finally studied the effect of random fluctuating
magnetic field. We have found that the parallel fluctuating field changes the singlet yield significantly,
but preserves the monotonicity of the singlet profile perfectly, and even can enhance the direction sensitivity.
Oppositely, the vertical fluctuating field destroys the monotonicity, and disrupts the orientation.
\begin{acknowledgments}
This work was supported by the National Natural Science
Foundation of China (Grants No. 11274043, 11075013, and 11005008).
\end{acknowledgments}

\bibliography{SFDM5BIBR2}

\begin{thebibliography}{}
\bibitem{quantum biology1} M. Arndt, T. Juffmann, and V. Vedral, {HFSP J.} \textbf{3}, 386 (2009).
\bibitem{quantum biology2} P. Ball, {Nature (London)} \textbf{474}, 272 (2011).
\bibitem{Engel}G. S. Engel, T. R. Calhoun, E. L. Read, T.-K. Ahn, T. Man\v{c}al, Y.-C. Cheng, R. E. Blankenship,
  and G. R. Fleming, {Nature (London)} \textbf{446}, 782 (2007).
\bibitem{Sarovar} M. Sarovar, A. Ishizaki, G. R. Fleming, and K. B. Whaley, {Nature Phys.} \textbf{6}, 462 (2010).
\bibitem{Collini} E. Collini and G. D. Scholes, {Science} \textbf{323}, 369 (2009).
\bibitem{Plenio} M. B. Plenio and S. F. Huelga, {New J. Phys.} \textbf{10}, 113019 (2008).
\bibitem{Mohseni} M. Mohseni, P. Rebentrost, S. Lloyd, and A. Aspuru-Guzik, {J. Chem. Phys.} \textbf{129}, 174106 (2008).
\bibitem{Briggs} John S. Briggs, and Alexander Eisfeld, {Phys. Rev. E} \textbf{83} 051911 (2011).
\bibitem{Ringsmuth} A. K. Ringsmuth, G. J. Milburn and T. M. Stace, {Nature Phys.} \textbf{8}, 562 (2012).
\bibitem{Nalbach} P. Nalbach, I. Pugliesi, H. Langhals, and M. Thorwart, {Phys. Rev. Lett.} \textbf{108}, 218302 (2012).
\bibitem{Lloyd} S. Lloyd, {Nature Phys.} \textbf{5}, 164 (2009).
\bibitem{Turin} L. Turin, {J. Theor. Biol.} \textbf{216}, 367 (2002).
\bibitem{Brookes} J. C. Brookes, F. Hartoutsiou, A. P. Horsfield, and A. M. Stoneham, {Phys. Rev. Lett.} \textbf{98}, 038101 (2007).
\bibitem{Harkins} T. T. Harkins and C. B. Grissom, {Science} \textbf{263}, 958 (1994).
\bibitem{Canfield} J. M. Canfield, R. L. Belfordoe, and P. G. Debrunner, {Mol. Phys.} \textbf{89}, 889 (1996).
\bibitem{Kominis} I. K. Kominis, {Phys. Rev. E} \textbf{80}, 056115 (2009).
\bibitem{Jianming} Jianming Cai, Gian Giacomo Guerreschi, and Hans J. Briegel, {Phys. Rev. Lett.} \textbf{104}, 220502 (2010).
\bibitem{Jones} J. A. Jones, P. J. Hore, {Chem. Phys.  lett.} \textbf{488}, 90 (2010).
\bibitem{Gauger} Erik M. Gauger, Elisabeth Rieper, John J. L. Morton, Simon C. Benjamin, and Vlatko Vedral, {Phys. Rev. Lett.}
      \textbf{106}, 040503 (2011).
\bibitem{Gauger2} Erik M. Gauger and Simon C. Benjamin, Phys. Rev. Lett. \textbf{110}, 178901 (2013).
\bibitem{Caijianming} Jianming Cai, {Phys. Rev. Lett.} \textbf{106}, 100501 (2011).
\bibitem{CYCai} C. Y. Cai, Qing Ai, H. T. Quan, and C. P. Sun, {Phys. Rev. A} \textbf{85}, 022315 (2012).
\bibitem{Cai} Jianming Cai, Filippo Caruso, and Martin B. Plenio, {Phys. Rev. A} \textbf{85}, 040304(R) (2012).
\bibitem{Jayendra} Jayendra N. Bandyopadhyay, Tomasz Paterek, and Dagomir Kaszlikowski, {Phys. Rev. Lett.} \textbf{109}, 110502 (2012).
\bibitem{Hogben} Hannah J. Hogben, Till Biskup, and P. J. Hore, {Phys. Rev. Lett.} \textbf{109}, 220501 (2012).
\bibitem{Stoneham} A. Marshall Stoneham, Erik M. Gauger, Kyriakos Porfyrakis, Simon C. Benjamin, and Brendon W. Lovett, {Biophysical Journal} \textbf{102}, 961 (2012).
\bibitem{Mouritsen} H. Mouritsen, and T. Ritz, {Current Opinion in Neurobiology} \textbf{15}, 406 (2005).
\bibitem{Johnsen} Johnsen, S., and K. J. Lohmann, {Physics Today} \textbf{61}, 29 (2008).
\bibitem{Ritz} T. Ritz, M. Ahmad, H. Mouritsen, R. Wiltschko, and W. Wiltschko, {Journal of The Royal
    Society Interface} \textbf{7}, S135 (2010).
\bibitem{Steiner} Steiner UE, Ulrich T, {Chem Rev.} \textbf{89}, 51 (1989).
\bibitem{Wiltschko} R. Wiltschko and W. Wiltschko, {BioEssays} \textbf{28}, 157 (2006).
\bibitem{Lohmann} S. Johnsen and K. J. Lohmann, {Nat. Rev. Neurosci.} \textbf{6}, 703 (2005).
\bibitem{Maeda} Kiminori Maeda, Kevin B. Henbest, Filippo Cintolesi, Ilya Kuprov, Christopher T. Rodgers, Paul A. Liddell,
    Devens Gust, Christiane R. Timmel, and P. J. Hore, {Nature (London)} \textbf{453}, 387 (2008).
\bibitem{Rodgers} C. T. Rodgers and P. J. Hore, {Proc. Natl. Acad. Sci. U.S.A.} \textbf{106}, 353 (2009).
\bibitem{Schulten} K. Schulten, C. E. Swenberg, and A. Weller, {Z. Phys. Chem.} \textbf{NF111}, 1 (1978).
\bibitem{Ritz2000} T. Ritz, S. Adem, and K. Schulten, {Biophys. J.} \textbf{78}, 707 (2000).
\bibitem{Wiltschko1972} W. Wiltschko and R. Wiltschko, {Science} \textbf{176}, 62 (1972).
\bibitem{Wiltschko1978} W. Wiltschko, in Animal Migration, Navigation, and Homing, edited by K. Schmidt-Koenig and W. T. Keeton
      (Springer, New York, 1978), p. 302.
\bibitem{Wiltschko2006} W. Wiltschko, K. Stapput, P. Thalau, and R. Wiltschko, {Naturwissenschaften} \textbf{93}, 300 (2006).
\bibitem{Ritz2004} Thorsten Ritz, Peter Thalau, John B. Phillips, Roswitha Wiltschko,
    and Wolfgang Wiltschko, {Nature (London)} \textbf{429}, 177 (2004).
\bibitem{Thalau} P. Thalau, T. Ritz, K. Stapput, R. Wiltschko, and W. Wiltschko, {Naturwissenschaften} \textbf{92}, 86 (2005).
\bibitem{Ritz2009} Thorsten Ritz, Roswitha Wiltschko, P. J. Hore, Christopher T. Rodgers, Katrin Stapput, Peter Thalau,
     Christiane R. Timmel, and Wolfgang Wiltschko, {Biophys. J.} \textbf{96}, 3451 (2009).
\bibitem{Dellis} A.T. Dellis, I.K. Kominis, {BioSystems} \textbf{107}, 153 (2012).
\bibitem{Kavokin} K.V. Kavokin, {Bioelectromagnetics} \textbf{30}, 402 (2009).
\bibitem{Tiersch} Markus Tiersch and Hans J. Briegel, {Philosophical Transactions of the Royal Society A} \textbf{370}, 4517 (2012).
\bibitem{Walters} Zachary B. Walters, arXiv:1208.2558 [physics. bio-ph] (2013).
\bibitem{Morton1} J. J. L. Morton, Alexei M. Tyryshkin, Arzhang Ardavan, Kyriakos Porfyrakis, S. A. Lyon, and G. Andrew D. Briggs,
        J. Chem. Phys. \textbf{124}, 014508 (2006).
\bibitem{Morton2} John J. L. Morton, Alexei M. Tyryshkin, Arzhang Ardavan, Kyriakos Porfyrakis, S. A. Lyon, and G. Andrew D. Briggs,
        Phys. Rev. B \textbf{76}, 085418 (2007).
\bibitem{Morton3} Richard M. Brown, Yasuhiro Ito, Jamie H. Warner, Arzhang Ardavan, Hisanori Shinohara, G. Andrew D. Briggs, and John J. L. Morton,
        Phys. Rev. B \textbf{82}, 033410 (2010).
\bibitem{Johnson} A. Johnson, J. Petta, J. Taylor, A. Yacoby, M. Lukin, C. Marcus, M. Hanson,
  and A. Gossard, {Nature (London)} \textbf{435}, 925 (2005).
\bibitem{Petta} J. Petta, A. Johnson, J. Taylor, E. Laird, A. Yacoby, M. Lukin, C. Marcus,
     M. Hanson, and A. Gossard, {Science} \textbf{309}, 2180 (2005).
\bibitem{Loreti} F. N. Loreti and A. B. Balantekin, {Phys. Rev. D} \textbf{50}, 4762 (1994).
\bibitem{Wang} Z. S. Wang, {Int. J. Theor. Phys.} \textbf{48}, 2353 (2009).
\bibitem{Wolf} Leonard Mandel and Emil Wolf, \emph{Optical coherence and quantum optics}, (Cambridge University Press, 2001).

\end{thebibliography}

\end{document}